\documentclass[12pt]{article}
\usepackage{amsfonts}
\font\twmsbm=msbm10 scaled 1200
\font\nmsbm=msbm9
\newfam\bbold
\textfont\bbold=\twmsbm
\scriptfont\bbold=\nmsbm
\scriptscriptfont\bbold=\nmsbm

\font\twscr=rsfs10 scaled 1200
\font\nscr=rsfs10
\newfam\script
\textfont\script=\twscr
\scriptfont\script=\nscr
\scriptscriptfont\script=\nscr

\newcommand{\be}{\begin{equation}}
\newcommand{\ee}{\end{equation}}
\newcommand{\bean}{\begin{eqnarray*}}
\newcommand{\eean}{\end{eqnarray*}}
\newcommand{\bea}{\begin{eqnarray}}
\newcommand{\eea}{\end{eqnarray}}

\newcommand{\const}{{\rm const}}
\newcommand{\tr}{{\rm tr}}
\newcommand{\rank}{{\rm rank}}

\newcommand{\p}{\partial}
\newcommand{\sq}{\sqrt{|g|}}

\title{String fluids and membrane media}
\author{M. G. Ivanov\thanks{e-mail: mgi@mi.ras.ru}\\
{\small\em Moscow Institute of Physics and Technology,}\\
{\small\em  141\,700, Dolgoprudnyi, Moscow Region, Russia }}
\date{}

\begin{document}
\maketitle

\abstract{
  String (membrane) theory could be considered
 as degenerate case of relativistic continuous media
 theory.
  The paper presents models of media,
 which are continuous distributions of interacting
 membranes, strings or particles.}

\section{Introduction (Kinematics)}
   To describe the motion of elastic medium one has
 to specify the motion of every {\em element of medium}.
   In the present paper {\em elements of medium} are
 particles, strings or membranes.
   To specify their motion one has to describe their
 world lines or world surfaces.

   Here we suppose, that world surfaces (lines) do not
 intersect each other, and every point of space-time
 $\bf M$ belongs to some world surface (line).

   Let world surfaces are numerated by parameters
 $\phi=(\phi^1,\dots,\phi^n)$ (coordinates across world surface).
   ${\bf V}_\phi$ is world surface number $\phi$.
   Let points of world surfaces are numerated by parameters
 $\xi=(\xi^1,\dots,\xi^{D-n})$ (coordinates along world surface).

   To describe the motion of elastic medium one can use
 explicit or implicit form
$$
   {\bf V}_\phi=\lbrace X\in{\bf M}|X=x(\xi,\phi)\rbrace,
\quad{\rm or}\quad
   {\bf V}_\phi=\lbrace X\in{\bf M}|\varphi(X)=\phi\rbrace.
$$
 I.e. space-time coordinates are functions $x=(x^0,\dots,x^{D-1})$
 of $\xi$ and $\phi$,
 or world surface ${\bf V}_\phi$ is intersection of level surfaces
 for $n$ space-time scalar fields
 $\varphi=(\varphi^1,\dots,\varphi^n)$.

   Explicit form is standard in relativistic string (membrane) theory
  [1--4].

   Implicit form was used in some papers on relativistic string (membrane)
 theory
 \cite{hos1,morris1,bf1,MGI-DAN,MGI-pbr,MGI-GRG},
 which deal with continuous distributions
 of strings (membranes).
   The aim of the present paper is to relate this field theory approach
 to relativistic elastic media mechanics.

   Implicit form allows to build differential form
 $J=f(\varphi)d\varphi^1\wedge\dots\wedge d\varphi^n$,
 which represents density of membranes.
   Some theories use other parametrizations $J=dI$ of form $J$
 (see \cite{GG} and references in these papers).
   These theories are discussed in Section \ref{I-phi-disc} of present paper.

   Kinematics of elastic media is described (in implicit form) by map
$
 \varphi:\>{\bf M}\to{\bf F},
$
 where
 $\bf M$ is pseudorimannian manifold  (space-time), $\dim{\bf M}=D$,
 $X^M$ are coordinates on $\bf M$ ({\em Euler coordinates}),
 $g_{MN}$ is metric;
 $\bf F$ is manifold, $\dim{\bf F}=n<D$,
 $\phi^\alpha$  are coordinates on $\bf F$ ({\em Lagrange coordinates}).

   World surface ${\bf V}_\phi$ is defined by means of
 inverse image ${\bf V}_\phi=\varphi^{-1}(\phi)$.

   Points of $\bf F$ represents elements of elastic medium.
   In further discussion we refer space $\bf F$ as {\em medium}.
   Now we do not introduce any additional structure at manifold $\bf F$,
 but later we have to add some structure on $\bf F$ to build
 appropriate action.

   In the kinematics we construct here gradients $d\varphi^\alpha$
 have to be space- or light- like.
   To study a causal structure of the theory one has
 to investigate the properties of energy-momentum tensor.

\section{Action\label{Sect-act-media}}

   To construct action for elastic medium we impose
 restrictions
 1) action does not involve higher derivatives of $\varphi$,
 2) field $\varphi$ is minimally coupled to metric,
 3) action does not involve other fields at $\bf M$,
 4) action is invariant under change of coordinates on $\bf M$
  (Euler coordinates),
 5) action is invariant under change of coordinates on $\bf F$
  (Lagrange coordinates).

   Items 1, 2  and 3 force us to write action in the following form
\be
  S_\phi=\int d^DX\sq~L_\phi(\varphi^\alpha,\p_M\varphi^\alpha,g_{MN}).
\ee

   Item 4 require Lagrangian $L_\phi$ to be scalar with respect
 to change of coordinates on $\bf M$.
   Under coordinate transformation on $\bf M$ fields
 $\varphi^\alpha$ are scalars, metric $g_{MN}$ is tensor,
 and gradients $\p_M\varphi^\alpha$ are covectors.
   Any scalar constructed from these fields is function
 of $\varphi^\alpha$ and scalar products of covectors
 $\p_M\varphi^\alpha$
\be
  {\cal G}^{\alpha\beta}=g^{MN}\,\p_M\varphi^\alpha\,\p_N\varphi^\beta.
\ee
   So, $L_\phi=L_\phi(\varphi^\alpha,{\cal G}^{\alpha\beta})$.

   Item 5 require Lagrangian $L_\phi$ to be scalar with respect
 to change of coordinates on $\bf F$.
   With respect to change of coordinates on $\bf F$
 matrix ${\cal G}^{\alpha\beta}$ is tensor
 (induced inverse metric at $\bf F$). 
   We are unable to build any nontrivial scalar from ${\cal G}^{\alpha\beta}$,
 so some additional parameters, which behave under change of coordinates
 on $\bf F$ as tensor fields, are necessary.
   These additional tensor fields specify the structure at $\bf F$,
 which was mentioned in previous section.
   From the point of view of field theory these tensors are functional
 parameters of action,
 from the point of view of elastic media mechanics they describes
 the medium (just in previous section we assign name {\em``medium''} to
 auxiliary space $\bf F$).

   Let matrix ${\cal G}^{\alpha\beta}$ is nondegenerate, i.e.
 $\det{\cal G}^{\alpha\beta}\not=0$
 ($\rank\,{\p_M\varphi^\alpha}=n$).
   Nondegeneracy condition means, that inverse image $\varphi^{-1}(\phi)$
 of point $\phi\in{\bf F}$ is (locally) a surface of dimension $D-n$.
   Medium we consider consist of such membranes, numerated by
 points of space $\bf F$.

   One can introduce inverse matrix ${\cal G}_{\alpha\beta}$ (metric)
 and volume form $\omega_{\bf F}$
\be
  {\cal G}^{\alpha\beta}{\cal G}_{\beta\gamma}=\delta^\alpha_\gamma,
 \quad
   {\cal G}=\det{\cal G}_{\alpha\beta}
          =\left(\det{\cal G}^{\alpha\beta}\right)^{-1},
 \quad
  \omega_{\bf F}=\sqrt{|{\cal G}|}
  \>d\phi^1\wedge\dots\wedge\,d\phi^n.
\label{omegaF}
\ee
   Tensors ${\cal G}_{\alpha\beta}$ and ${\cal G}^{\alpha\beta}$
 are assumed to be used for raising and lowering
 of indices at space $\bf F$ (Greek indices).

   Coordinate system with $D-n$ coordinates
 specified by relation $x^\alpha=\varphi^\alpha$
 is {\em attendant coordinate system}.
   Energy density in attendant coordinates is $\rho=-L$.

\section{Energy-momentum and stress tensors}

  By variation of metric, or variation of $\p_K\varphi^\alpha$
 one can find energy-momentum tensor
 (i.e. general relativity definition of energy-momentum
 tensor and mechanical definition produce the same
 expression)
\be
  T_{MN}
    =g_{MN}L_\phi
    -2\frac{\p L_\phi}{\p{\cal G}^{\alpha\beta}}
     \,\p_M\varphi^\alpha\,\p_N\varphi^\beta
    =g_{MK}
     \left(\delta^K_N L_\phi
    -\frac{\p L_\phi}{\p(\p_K\varphi^\alpha)}
     \,\p_N\varphi^\alpha
     \right).
\ee
\be
  \nabla_KT^K{}_N=\partial_N\varphi^\alpha
  \left(
    \frac{\p L_\phi}{\p\varphi^\alpha}
   -\nabla_K\frac{\p L_\phi}{\p(\p_K\varphi^\alpha)}
  \right)
  =\partial_N\varphi^\alpha\>\frac1{\sqrt{|g|}}
   \frac{\delta S_\phi}{\delta\varphi^\alpha}.
\label{zn-soxr-en-imp}
\ee
  So, if $\rank \p_N\varphi^\alpha=n$
 then field equations
 $\frac{\delta S_\phi}{\delta\varphi^\alpha}=0$
 are equivalent to energy-momentum conservation 
 $\nabla_MT^M{}_N=0$.
  
  Contravariant energy-momentum tensor $T^{MN}$
 could be projected by $\varphi_*$ from space-time $\bf M$
 to medium $\bf F$.
  New medium tensor $S^{\mu\nu}$ we refer as
 {\em stress tensor}\footnote{Under $(-,+,+,+)$ space-time
 signature we use here, tensor (\ref{Smunu}) has opposite
 sign with respect to standard stress tensor introduced
 in mechanics.}
\be
  S^{\mu\nu}=T^{MN}\>\p_M\varphi^\mu\>\p_N\varphi^\nu
 ={\cal G}^{\mu\nu}\,L_\phi
 -2\frac{\p L_\phi}{\p{\cal G}^{\alpha\beta}}
  \>{\cal G}^{\alpha\mu}\>{\cal G}^{\beta\nu}.
\label{Smunu}
\ee
   It looks like energy-momentum tensor formula with
 $g^{MN}$ replaced by ${\cal G}^{\alpha\beta}$.

\section{Perfect membrane fluid}

   It was stated in
 section \ref{Sect-act-media}
 that to build Lagrangian, which is scalar with respect
 to $\bf M$ and $\bf F$, one has to specify at $\bf F$ some
 tensor fields, which are parameters of action describing
 the medium.
   The simplest choice is to introduce at $\bf F$ a volume
 form $\Omega_{\bf F}=f(\phi)\>d\phi^1\wedge\dots\wedge\,d\phi^n$.
   Using volume forms $\Omega_{\bf F}$ and $\omega_{\bf F}$
 one can build the scalar, which is ratio of volumes
 $\frac1{n!}\,
  \Omega_{\bf F}^{\alpha_1\dots\alpha_n}
  \omega_{{\bf F}\,\alpha_1\dots\alpha_n}
 =f(\varphi)/\sqrt{|\cal G|}$.

   Lagrangian $L_\phi$ is a function $L_{\rm fluid}$
 of $\varphi$ and scalar $f(\varphi)/\sqrt{|\cal G|}$.
\be
  S_{\rm fluid}=\int d^Dx\sqrt{|g|}\>
   L_{\rm fluid}
   \left(\varphi,f(\varphi)\sqrt{\det{\cal G}^{\alpha\beta}}\right)
\label{Sfluid}
\ee
   Action (\ref{Sfluid})
 with $n=D-1$ describes fluid or gas (with no dissipation).
   If $n<D-1$ one has analogue of fluid, which consists of
 membranes with nonintersecting world surfaces.

   This action was considered in paper \cite{MGI-GRG}.

   Equations of motion for simple media are consequence
 of energy-momentum conservation, so to check correspondence
 of model (\ref{Sfluid}) to fluid one can compare  just
 energy-momentum tensors
\be
  T_{MN}=P_{MN}\,L_{\rm fluid}+\bar P_{MN}\,P_{\rm fluid}.
\label{T_MNfluid}
\ee
   Here $P_{MN}$ is projector to world surface (surface $\varphi=\const$), 
 $\bar P_{MN}=g_{MN}-P_{MN}
  ={\cal G_{\alpha\beta}}\,\p_M\varphi^\alpha\,\p_N\varphi^\beta$
 is projector to directions orthogonal to world surface.
   Pressure (in attendant coordinate system)
 is derived from Lagrangian by following ``Legendre transformation''
 (prime means derivative by $f(\varphi)/\sqrt{|\cal G|}$)
\be
  P_{\rm fluid}=L_{\rm fluid}
 -L'_{\rm fluid}\,f\,\sqrt{\det{\cal G}^{\alpha\beta}}.
\label{equation_of_state}
\ee

   If pressure $P_{\rm fluid}$ vanish (linear $L_{\rm fluid}$), then
 one has gas of membranes with zero pressure, i.e. each
 world surface $\varphi=\const$ behaves as free membrane.
 (Equivalence of equations of motion in this case with
 equations of motion of free membrane was demonstrated in
 papers \cite{MGI-DAN}.)

 {\bf Example:} fluid with linear equation of state
 $P_{\rm lin.fluid}=k\,\rho_{\rm lin.fluid}$
 is described by Lagranigian
 $L_{\rm lin.fluid}=-\left(\det{\cal G}^{\alpha\beta}\right)^{\frac{1+k}2}$.

\section{Membrane fluids and nonlinear electrodynamics-type theories
  \label{I-phi-disc}}

   One can write action (\ref{Sfluid}) in terms of closed
 form $J$ (``current'')
\be
 J=\varphi^*\Omega_{\bf F}=f(\varphi)\>d\varphi^1\wedge\dots\wedge\,d\varphi^n.
\label{J-dvarphi}
\ee
$$
  \|J\|=\sqrt{\frac1{n!}J_{M_1\dots M_n}J^{M_1\dots M_n}}
  =f(\varphi)\sqrt{\det{\cal G}^{\alpha\beta}}.
$$
$$
  S_{\rm fluid}[\varphi]=\int d^Dx\sqrt{|g|}\>
   L_{\rm fluid}
   \left(\varphi,\|J\|\right).
$$

   If fluid is homogeneous, then Lagrangian is independent
 on $\varphi$, i.e. $L_{\rm h.fluid}(\|J\|)$.
   So,
\be
  S_{\rm h.fluid}[\varphi]
  =\int d^Dx\sqrt{|g|}\>L_{\rm h.fluid}\left(\|J\|\right).
\label{SHfluid}
\ee

  One can use other definition of $J$
\be
  J=dI.
\label{J-dI}
\ee
   Any $J$ defined by (\ref{J-dvarphi}) could be written
 in form (\ref{J-dI}).

  Definition (\ref{J-dI}) allows to build 
 nonlinear electrodynamics-type theory with
 action
\be
  S_{\rm nonlin.}[I]
  =\int d^Dx\sqrt{|g|}\>L_{\rm h.fluid}\left(\|J\|\right).
\label{nonlin}
\ee
  Theories of this type are also called ``string fluids'' \cite{GG}.

  For both actions (\ref{SHfluid}), (\ref{nonlin}) the only
 physical field is $J$.
  E.g. both energy-momentum tensors are written in terms of
 field $J$ only by the same formula.
  Fields $I$ and $\varphi$ play the role of ``potentials'',
 i.e. they are used to derive field equations, which could
 be written in terms of $J$ only.\footnote{
  Field equations for (\ref{SHfluid}) has the following two forms
 (see \cite{MGI-GRG})
$$
  \frac{\delta S_{\rm h.fluid}}{\delta\varphi^\alpha}=0
 \Leftrightarrow
  \frac{\delta S_{\rm h.fluid}}{\delta\varphi^\alpha}\p_M\varphi^\alpha=0.
$$
  The last one could be written in terms of $J$ only.}

  So, both theories look very similar, but
 nonlinear electrodynamics-type theory
 is different from membrane fluid theory
 (even if the function $L_{\rm h.fluid}$ is the same).
  If field $J$ could be represented in form (\ref{J-dvarphi})
 (i.e. in both forms (\ref{J-dvarphi}) and (\ref{J-dI})), then
\be
   \frac{\delta S_{\rm h.fluid}}{\delta\varphi^\alpha}
  =\frac1{(n-1)!}\frac{\delta S_{\rm nonlin.}}{\delta I_{M_1\dots M_{n-1}}}
   \varepsilon_{\alpha\alpha_1\dots\alpha_{n-1}}
   f(\varphi)
   \p_{M_1}\varphi^{\alpha_1}\dots\p_{M_{n-1}}\varphi^{\alpha_{n-1}},
\ee
 or (the other form of same relation)
$$
   \frac{\delta S_{\rm h.fluid}}{\delta\varphi^\alpha}\p_M\varphi^\alpha
 =\frac1{(n-1)!}\frac{\delta S_{\rm nonlin.}}{\delta I_{M_1\dots M_{n-1}}}
  J_{MM_1\dots M_{n-1}},
$$
 here
\be
  \frac{\delta S_{\rm nonlin.}}{\delta I_{M_1\dots M_{n-1}}}
 =-\p_{M}
  \left(\sqrt{|g|}\frac{\p L_{\rm h.fluid}}{\p J_{MM_1\dots M_{n-1}}}\right).
\ee

   I.e. definition (\ref{J-dI}) admits wider set of fields $J$,
 and generates stricter field equations.

   Fields $J$, which could be represented in form (\ref{J-dvarphi})
 and satisfy field equations
 $\frac{\delta S_{\rm nonlin.}}{\delta I_{M_1\dots M_{n-1}}}=0$
 are solutions for both theories.
   Some fields $J$, which could be represented in form (\ref{J-dvarphi})
 and satisfy field equations
 $\frac{\delta S_{\rm h.fluid}}{\delta\varphi^\alpha}=0$
 are {\em not} solutions of field equations
 $\frac{\delta S_{\rm nonlin.}}{\delta I_{M_1\dots M_{n-1}}}=0$.
    Similarly some fields $J$, which satisfy field equations
 $\frac{\delta S_{\rm nonlin.}}{\delta I_{M_1\dots M_{n-1}}}=0$
 could not be represented in form (\ref{J-dvarphi})
 (these fields do not represent membrane fluids).

\section{Elastic membrane media}

   Elastic medium in contrast to fluid tends to preserve not
 only volume, but also form, i.e. distance between close
 points, which determined by {\em metric}.

   In classical mechanics it means existence of certain
 ``nondeformed metric'', deviation from which is deformation.
   So, at space $\bf F$ one has to define metric
 ${\cal G}^0_{\alpha\beta}$, which does not depend on
 map $\varphi$ and describes medium itself.
   Metric ${\cal G}^0_{\alpha\beta}$ specifies
 {\em nondeformed state}.

   {\em Nondeformed state} is defined as state
 with stress tensor $S^{\mu\nu}$ vanishing and
 zero external forces (including gravity).

   In the framework of general relativity defining of
 such state looks impossible, but stress tensor
 $S^{\mu\nu}$ is defined (\ref{Smunu})
 in terms of internal geometry of space $\bf F$.
   ``Nondeformed state'' we are looking for is
 metric at $\bf F$, so one can formulate the
 following problem
\be
  \left.S^{\mu\nu}\right|_{{\cal G}_{\alpha\beta}={\cal G}^0_{\alpha\beta}}=0.
\label{SmunuG0}
\ee
   Metric ${\cal G}^0_{\alpha\beta}$, which is solution of (\ref{SmunuG0})
 is nondeformed state.

   One could refer medium as {\em elastic medium}
 iff equations (\ref{SmunuG0}) possess discrete
 set of solutions.

   Strain tensor for elastic medium is defined as
 $\epsilon_{\mu\nu}=\frac12\left({\cal G}_{\mu\nu}-{\cal G}^0_{\mu\nu}\right)$.

\subsection{Isotropic elastic membrane media}

   Isotropic medium is characterised by metric $g_{{\bf F}\>\alpha\beta}$
 (metric $g_{{\bf F}\>\alpha\beta}$ could be different
 from ${\cal G}_{\alpha\beta}^0$)
 and some scalar fields at $\bf F$.

   Let us introduce the following notation
 $\tilde{\cal G}^\alpha{}_\gamma
  ={\cal G}^{\alpha\beta}g_{{\bf F}\>\beta\gamma}$.

   By using two $n$-dimensional metrics
 ($g_{{\bf F}\>\alpha\beta}$ and ${\cal G}_{\alpha\beta}$)
 one can build $n$ independent scalars.
   E.g. one can consider roots or coefficients of secular equation
\be
  \det(\tilde{\cal G}^\alpha{}_\gamma-\lambda\,\delta^\alpha{}_\gamma)
 =\sum_{k=0}^n (-\lambda)^{n-k}\>f_k=0,\qquad
  f_k=
  \tilde{\cal G}^{[\alpha_1}{}_{\alpha_1}
  \dots
  \tilde{\cal G}^{\alpha_k]}{}_{\alpha_k}
\label{f_k}
\ee
$$
  f_0=1,\quad
  f_1=\tilde{\cal G}^\alpha{}_\alpha=\tr\tilde{\cal G},\quad
  f_2=\frac12\left((\tr\tilde{\cal G})^2-\tr(\tilde{\cal G}^2)\right),\quad
  f_n=\det\tilde{\cal G}^\alpha{}_\gamma=\frac{g_{\bf F}}{\cal G}.
$$

   Lagrangian of isotropic elastic membrane medium
 is
 function of $f_k$ and some scalars at $\bf F$
\be
   S_{\rm iso.}=\int\limits_{\bf M}d^Dx\,\sqrt{|g|}
  L_{\rm iso.}(\varphi,f_1,\dots,f_n).
\ee
   If $L_{\rm iso.}$ depends on $\varphi$ and $f_n$ only,
 then it describes again perfect membrane fluid.

\subsection{Linear isotropic membrane media}

   To consider equations of motion and energy-momentum tensor
 in linear order on strain tensor $\epsilon_{\mu\nu}$,
 one need scalars $f_k$ (\ref{f_k}) up to second order on $\epsilon_{\mu\nu}$.
   Under this accuracy any two scalars $f_k$ could be used to express
 all other scalars $f_k$.
   So, to construct action (under this level of accuracy)
 one can choose any two scalars $f_{k_1}$ and $f_{k_2}$.
   It is just the matter of convenience.

   It is convenient to use $f_n=\det\tilde{\cal G}$,
 to simplify comparison with perfect fluid.
   Another scalar, which is natural to use is
$f_1=\tr\tilde{\cal G}
    =g^{MN}\>\p_M\varphi^\alpha\>\p_N\varphi^\beta\>g_{{\bf F}\,\alpha\beta}$.
   Its form reproduces standard $\sigma$-model Lagrangian
 for set of scalar fields.

   In linear approximation isotropic elastic membrane medium
 is described by two elastic moduli $\mu$, $\lambda$,
 density in nondeformed state $\rho_0$ and
 metric of non-deformed state
 ${\cal G}_{\alpha\beta}^0=g_{{\bf F}\,\alpha\beta}$.
   E.g. the possible Lagranian is
\be
  L_{\rm lin.}=
  -\frac{(\rho_0-\mu)^2}{\rho_0+\lambda}
   \left(f_n^{\frac12\left(\frac{\rho_0+\lambda}{\rho_0-\mu}\right)}-1\right)
  -\frac\mu2\,(f_1-n)-\rho_0.
\label{Llin}
\ee


\subsection{Perturbations of linear membrane media}

   Lagrangian (\ref{Llin}) requires gradients $d\varphi^\alpha$
 to be space-like.
   It is natural to study propagation of perturbations
 representing fields $\varphi^\alpha$ in the
 form $\varphi^\alpha=\zeta^\alpha+\chi^\alpha$.
   Here $\zeta^\alpha$ is a solution of equations
 of motion generated by
 Lagranigian (\ref{Llin}), and
 $\chi^\alpha$ is small perturbation
 ({\em displacement vector}).
   Gradients $d\chi^\alpha$ are assumed to be
 time- or light- like.
   Expansion of Lagrangian (\ref{Llin}) in series
 on $\chi^\alpha$ allows to build perturbation theory.

   Let space-time is direct product of longitudinal  space $\bf V$
 with metric $\gamma_{mn}$ and medium space $\bf F$, i.e.
 $ds^2=\gamma_{mn}dz^mdz^n+g_{{\bf F}\,\alpha\beta}dx^\alpha dx^\beta$.
   Metric coefficients $\gamma_{mn}$,
 $g_{{\bf F}\,\alpha\beta}$, unperturbed
 density $\rho_0$ and elastic moduli
 $\lambda$ and $\mu$ are independent on coordinates.
   To lower and raise Greek and Latin indices we use
 matrices $g_{{\bf F}\,\alpha\beta}$ and $\gamma_{mn}$
 and inverse matrices.

   Fields $\zeta^\alpha=x^\alpha$ are solution of
 equations of motion, because energy-momentum tensor
 $T^{(0)}_{MN}=-\rho_0 P_{MN}$
 (here $P_{MN}$ is projector to surface $\zeta=\const$)
 obviously satisfies conservation law.

   Up to quadratic terms one has
\bea
  L_{\rm lin.}&\approx&
  -\rho_0\left(1+\p_\alpha\chi^\alpha
             +\frac12\p_m\chi^\alpha\,\p^m\chi_\alpha\right)-
\\
\nonumber
  {}&&-\frac{\rho_0+\lambda}2\,(\p_\alpha\chi^\alpha)^2
  -\frac\mu2\,\p_\alpha\chi^\gamma\,\p^\alpha\chi_\gamma
  +\frac{\rho_0-\mu}2\,\p_\alpha\chi^\gamma\,\p_\gamma\chi^\alpha.
\eea 
   Linearized equations of motion have the following form
\be
  \frac{\delta S_{\rm lin.}}{\delta\chi^\alpha}\approx
 -\rho_0\, \p_m\p^m \chi_\alpha
 +\mu\,\p_\mu\p^\mu \chi_\alpha
 +(\lambda+\mu)\,\p_\alpha(\p_\gamma\chi^\gamma).
\ee

   It is obvious that this equation
 admit transverse and longitudinal waves,
 which propagate along coordinates $x$
 with velocities $C_\bot$ and $C_{\|}$
 and perturbations, which propagate
 along coordinates $z$
 (iff $\dim{\bf V}>1$, i.e.
 iff the medium consists of strings or membranes, but not of particles)
 with speed of light
\be
   C_\bot=\sqrt{\frac\mu{\rho_0}},\qquad
   C_{\|}=\sqrt{\frac{\lambda+2\mu}{\rho_0}},\qquad
   C_{\bot'}=1.
\ee

   In the case $\dim{\bf V}=1$, one has the standard equations
 for displacement vector in homogeneous isotropic elastic medium.

\section{Conclusion}

   Consideration of linearized theory is first step
 of quantizing of initial model (\ref{Llin}).
   In the linearized theory quantization is just
 introducing of creation-annihilation operators
 for phonons.
   The next stage is perturbation theory, which have
 to take into account self-interaction of fields $\chi$,
 originated from high order term.

   The models considered admits various generalisations,
 e.g. one can easily introduce interactions with other fields.
   (Lagrangian for membrane media itself could be used
 as interaction term for scalar fields $\varphi^\alpha$.)

   Existence of volume form $\Omega_{\bf F}$ at space $\bf F$
 allows to introduce current $J=\,\varphi^*\Omega_{\bf F}$
 at space $\bf M$.
   Current $J$ satisfies ``kinematic conservation law'' $dJ=0$
 \cite{MGI-DAN}.
   E.g. if $\dim{\bf M}=\dim{\bf F}+1$, then vector $*J$
 is standard $D$-dimensional current density.

   To describe interaction one has just to include
 in action standard current-field interaction term,
 e.g. interaction with electromagnetic field described by
 $\int_{\bf M}J\wedge A$.

   The considered models of ``membrane media''
 reproduces some properties of strings and
 membranes, which are elements of medium.
   It reveals the close relation between modern string
 theory and relativistic continuous media theory.
   This relation has to become fruitful for both theories.

\subsection*{Acknowledgement}
  The author is grateful to I.V. Volovich and M.O. Katanaev.
  The work was partially supported by grants
 RFFI 02-01-01084, NSh-1542.2003.1.

\end{document}